\documentclass{article}

\usepackage[left=25mm,right=25mm]{geometry}
\usepackage{amsmath,amssymb}
\usepackage{bm}

\newtheorem{theorem}{Theorem}

\newtheorem{lemma}[theorem]{Lemma}

\newcommand{\bra}[1]{\langle #1 |}
\newcommand{\ket}[1]{| #1 \rangle}
\newcommand{\bracket}[2]{\langle #1 | #2 \rangle}
\newcommand{\ketbra}[2]{| #1 \rangle \langle #2 | }
\newcommand{\HA}{\mathop{\mathcal{H}}\nolimits} 

\begin{document}

\title{Solutions to the mean king's problem: \\higher-dimensional quantum error-correcting codes}
\author{Masakazu Yoshida, Toru Kuriyama, Jun Cheng \\
\small{Faculty of Science and Engineering, Doshisha University}}
\date{}
\maketitle

\begin{abstract}
Mean king's problem is a kind of quantum state discrimination problems. 
In the problem, we try to discriminate eigenstates 
of noncommutative observables with the help of classical delayed information. 
The problem has been investigated from the viewpoint of error detection and correction. 
We construct higher-dimensional quantum error-correcting codes 
against error corresponding to the noncommutative observables. 
Any code state of the codes provides a way to discriminate the eigenstates correctly with the classical delayed information. 
\end{abstract}


\section{Introduction}
Mean king's problem is 
a kind of quantum state discrimination  problems formulated by 
Vaidman, Aharanov, and Albert \cite{VAA87}. 
The problem is often told as a tale of a king and a physicist Alice. 
At first, Alice prepares a quantum bit (qubit)-system in an initial state. 
King performs a measurement with one of observables $\sigma_x$, $\sigma_y$, $\sigma_z$ and obtains an outcome. 
After the king's measurement, 
Alice performs a measurement and obtains an outcome. 
After the Alice's measurement, 
king reveals the observables he has measured. 
Then, she guesses the king's outcome by using her outcome and the classical delayed information from king. 
A solution to the problem is defined as a pair of the initial state and the Alice's measurement 
such that she can guess the king's outcome correctly.  
A solution has been shown by making use of a bipartite qubits-system in a Bell state \cite{VAA87}. 
Then, Alice keeps one of the qubits and king performs the measurement on the other one. 
Her measurement derived from Aharonov-Bergman-Lebowitz rule \cite{ABL_rule} is performed on the bipartite qubits-system. 

The mean king's problem is considered in several settings. 
Most naturally, 
it is considered that 
king employs measurements constructed from a complete set of mutually unbiased bases (MUBs) \cite{WF89,Ivanovic}. 
In the setting, 
when a bipartite system is prepared in a maximally entangled state, 
the existence of a solution to the problem depends on the existence of a complete set of orthogonal Latin squares \cite{HHH05}. 
Note that the complete sets of MUBs and orthogonal Latin squares exist in prime-power dimension. 
For arbitrary dimension, 
it has been shown that solutions always exist when Alice is allowed to employ a positive operator valued-measure (POVM) measurement \cite{KTO06}. 
Nonexistence of the solutions to the problem has been shown 
when Alice cannot prepare a bipartite system \cite{Aravind03,Kimura07}. 

By investigating the problem from the viewpoint of error detection and correction, 
a solution to the problem by using quantum error-correcting codes has been shown \cite{YKMIC15}. 
A quantum error-correcting code is defined as a subspace of a Hilbert space 
and a quantum state in the code is called a code state. 
Roughly speaking, correction of error on the state is realized with discrimination of added error states 
and performing an appropriate quantum operation effectively. 
Applying the context of quantum error-correcting codes to the problem, 
Alice prepares a code state, which is in a quantum error-correcting code against error corresponding to the king's measurements, as an initial state. 
Then, she can guess the king's outcome by discriminating error with the help of the classical delayed information from king. 
In the previous works, 
since the specific quantum state, e.g., the maximally entangled state, is considered for the initial state, 
it is not clear how large solution space is. 
On the other hand, any code state of the code is considered for the initial state in this method (we recall that the code is the subspace of the whole space). 
However, the general construction of such quantum error-correcting codes under any problem setting is not known. 

In this paper, 
we show {\it higher-dimensional} quantum error-correcting codes. 
Here, {\it higher-dimensional} means that the dimension of the code is greater than or equal to $2$. 
On a bipartite system, which consists of different dimensional local systems, 
and amultipartite system (this case is outside the purview of the setting as stated above), 
we provide constructions of such quantum codes based on 
some properties of a pair of an entangled state and a measurement. 
Then, Alice can guess the king's outcome correctly by using any code state of the codes as an initial state 
if the pair of the state and the measurement provides a solution to the problem. 
This implied that we can find more large solution space  in the context of the solution by using quantum error-correcting codes 
if the there exists the solution. 
We also show some examples of the higher-dimensional codes 
in the case that the king's measurements are performed on the qubit system. 

This paper is organized as follows. 
In Sec. 2, 
we review quantum error-correcting codes and 
the solution to the mean king's problem by using quantum error-correcting codes. 
In Sec. 3 and Sec. 4, 
we show the constructions of higher-dimensional quantum error-correcting codes 
in the bipartite system and  the multipartite system, respectively. 
We also show some examples of the constructed codes on the qubits system. 
Finally, in Sec. 5, we summarize this paper.

\section{Review of Solutions Using Quantum Error-Correcting Codes}
In this section, 
we review the basics of quantum error-correcting codes and 
introduce the solution to the mean king's problem 
using quantum error-correcting codes \cite{YKMIC15}. 

We regard $d$-dimensional Hilbert spaces in the same light as $d$-level quantum systems 
and call a $2$-level quantum system a quantum bit (qubit) system. 
In this paper, we treat a quantum operation described by 
a trace nonincreasing completely positive (CP) map 
as adding error to a quantum system $\cal H$. 
$\epsilon : {\cal S}(\cal H)\rightarrow {\cal L}(\cal H)$ is a 
trace nonincreasing if and only if there exist operators $(E_k)_k$ 
satisfying $\epsilon (\rho)=\sum_{k}E_k^{\dagger}\rho E_k$ for $\rho\in {\cal S}(\cal H)$ and 
$\sum_kE_k^{\dagger}E_k \leq \mathbb I$, 
where ${\cal S}(\cal H)$ and ${\cal L}(\cal H)$ are the sets of density operators and liner operators on $\cal H$, respectively. 
This representation of $\epsilon$ $(E_k)_k$ is called the Kraus representation. 
We identify a trace nonincreasing CP map $\epsilon$ with its Kraus representation $(E_k)_k$. 
Furthermore, we call $\epsilon$ (or $(E_k)_k$) an error in the context of quantum codes. 
A $d'$-dimensional subspace of a $d$-dimensional Hilbert space 
is called a $(d,d')$ quantum code 
and a unit vector in the subspace is called a code state. 
We omit the notation $(d,d')$ when the context is clear. 
A $(d,d')$ quantum code $C$ is called a $(d,d')$ quantum error-correcting code 
against an error $\epsilon$ 
if there exists a trace-preserving completely positive map $R$ such that 
$R(\epsilon(\rho)) \propto \rho$ 
holds for any $\rho \in S(C)$. 
Such $R$ is called a recovery. 
The general condition for the existence of quantum error-correcting codes was given by Knill and Laflamme \cite{Nieal Laframme}. 
Let $C$ be a $(d, d')$ quantum code and $(E_k)_k$ an error. 
There exists a recovery $R$ for $C$ to be a quantum error-correcting code against $(E_k)_k$ 
if and only if 
$P E_k^{\dagger} E_{k'} P = \alpha_{kk'} P $ 
holds, where $(\alpha_{kk'})_{k,k'}$ is a positive matrix 
and $P$ denotes the projection onto $C$.

We review the solution to the mean king's problem using quantum error correcting codes. 
The problem has been summarized as follows: 
\begin{enumerate}
\item 
Alice prepares a bipartite system ${\cal H}_A \otimes {\cal H}_K$ in an initial state, 
where ${\cal H}_A$ (resp. ${\cal H}_K$) is a $d_A$-dimensional (resp. $d_K$-dimensional) Hilbert space. 
\item 
King performs one of measurements $M^{(J)}$ ($J=1,2,\ldots ,n$) on ${\cal H}_K$, 
which are described by measurement operators
$(M_i^{(J)})_{i=1}^m$, and obtains an outcome $i$ 
\footnote{
Let $(M_i^{(J)})_i$ be a collection of measurement operator on a Hilbert space $\HA$. 
The postmeasurement state from the $\rho\in {\cal S}(\HA)$ is given by 
$M_i^{(J)}\rho {M_i^{(J)}}^\dagger / \mathrm{tr}M_i^{(J)}\rho {M_i^{(J)}}^\dagger$ 
and the probability to get the outcome $i$ is 
$\mathrm{tr}M_i^{(J)}\rho {M_i^{(J)}}^\dagger$. 
}. 
\item 
Alice performs a measurement described by a POVM ${\cal P}$ 
on the bipartite system and obtains an outcome $k$. 
\item 
King reveals the measurement type $J$ he has performed. 
\item 
Immediately, Alice guesses $i$ by using $k$ and $J$. 
\end{enumerate}
In the above setting, 
a solution to the problem is defined as 
a pair of the initial state and the Alice's measurement 
such that she can guess the king's outcome correctly. 
Then, the following theorem was given. 

\begin{theorem}\label{QKP_QECC}(Theorem 3 in Ref. \cite{YKMIC15})
Let $C \subset {\cal H}_A \otimes {\cal H}_K$ be a $(d_Ad_K,d')$ quantum code 
and $P$ the projection onto $C$. 
If there exist $l$ tuple operators on ${\cal H}_K$ 
$(L_k)_{k=1}^l$ with $\sum_k L_k^\dagger L_k \leq \mathbb I$ 
and nonempty index sets $X^{(J,i)} \subset \{1,2,\ldots l\}$ ($J=1,2,\ldots , n$ : $i=1,2\ldots ,m$) 
satisfying the following conditions: 
\begin{eqnarray}
&&X^{(J,i)} \cap X^{(J,i')} = \emptyset \hspace{0.3cm} \forall J,i,i',  \label{Thm_emp}\\
&&\mathbb{I} \otimes M_i^{(J)} = \sum_{k\in X^{(J,i)}} f^{(J,i)}_{k}\mathbb{I} \otimes L_k 
\hspace{0.2cm} \mathrm{on} \hspace{0.2cm} C \hspace{0.3cm} \forall J,i, \label{Thm_decom}\\
&&P (\mathbb{I} \otimes L_k)^{\dagger} (\mathbb{I} \otimes L_{k'}) P = \lambda_k \delta_{kk'} P \label{Thm_ortho}
\end{eqnarray}
for some $\lambda_k \geq 0$ and $f^{(J,i)}_{k}\in {\mathbb C}$. 
Then, 
\begin{itemize}
\item[(i)] 
Alice can guess the king's outcome correctly by using any code state in $C$ as an initial state, 
\item[(ii)] 
$C$ is a quantum error-correcting code against an error $(\mathbb{I} \otimes L_k)_k$. 
\end{itemize}
\end{theorem}

An outline of the proof of Theorem \ref{QKP_QECC} is as follows. 
(i) a subspace which contains a state after the king's measurement is 
uniquely determined by $(L_k)_{k\in X^{(J,i)}}$ from Eq. \eqref{Thm_decom} 
and such subspaces are orthogonal from Eq. \eqref{Thm_ortho}. 
$X^{(J,i)}$ which contains $k$ uniquely exists for $J$ from Eq. \eqref{Thm_emp}. 
Then, Alice can guess the king's outcome by using her outcome and $J$ 
when she performs a measurement to distinguish the subspaces 
(i.e., this measurement is described by a projection valued measure (PVM), 
which consists of the projections onto the subspaces).  
(ii) It is straightforward 
from  Eq. \eqref{Thm_ortho} and the Knill-Laflamme theorem \cite{Nieal Laframme} as stated above. 

Here, we give one points. 
Theorem \ref{QKP_QECC} says that 
there exists a solution to the problem if there exists 
a quantum error-correcting code against the error $(\mathbb{I} \otimes L_k)_k$ 
satisfying Eqs. \eqref{Thm_emp}, \eqref{Thm_decom}, 
and \eqref{Thm_ortho} for the king's measurement. 
However, the state-change caused by the error (quantum operation) $(\mathbb{I}\otimes L_k)_k$ 
deffer from the measurement process of the measurement operators $(M_i^{(J)})_i$. 
The state-change of $\ket{\psi}\in C$ against the error $(\mathbb{I} \otimes L_k)_k$ 
is described by $\ketbra{\psi}{\psi}\mapsto \sum_{k}(\mathbb{I}\otimes L_k )\ketbra{\psi}{\psi}(\mathbb{I}\otimes L_k )^{\dagger}=:\rho'$. 
The post-state is recovered by a map $R(\rho')\propto \ketbra{\psi}{\psi}$. 
At that time, error detection is not necessarily required. 
On the other hand, 
in the context of mean king's problem, 
the initial state $\ket{\psi}\in C$ is changed by 
a king's measurement described by measurement operators $(M_i^{(J)})_i$ 
with an outcome $i$: 
$\ketbra{\psi}{\psi}\mapsto (\mathbb I\otimes M_i^{(J)})\ketbra{\psi}{\psi}(\mathbb I\otimes M_i^{(J)})^\dagger / p_i =: \sigma$, 
where $p_i= \mathrm{tr}(\mathbb I\otimes M_i^{(J)})\ketbra{\psi}{\psi}(\mathbb I\otimes M_i^{(J)})^\dagger$ is the probability to get the outcome $i$. 
Then, we observe 
$\sigma= (\sum_{k\in{X^{(J,i)}}}\mathbb I\otimes L_k)\ketbra{\psi}{\psi}
(\sum_{k\in{X^{(J,i)}}}\mathbb I\otimes L_k)^\dagger / p_i$. 
Therefore, Alice can guess the king's outcome $i$ with 
an outcome of the measurement to distinguish $(\mathbb I\otimes L_k\ket{\psi})_k$ and $J$. 

A ``reverse'' statement of Theorem \ref{QKP_QECC} was given. 
Let $\HA_A=\HA_K := \HA$ ($\dim \HA = d$) and an entangled state (in the form of the Schmidt decomposition): 
\begin{equation}\label{eq:Sch}
\ket{\Psi_{\bm \eta}} := \sum_{j=0}^{d-1} \eta_j \ket{\psi_j} \otimes \ket{\phi_j} 
\hspace{0.2cm} \eta_j > 0, \sum_{j=0}^{d-1}\eta_j^2=1 
\end{equation}
be prepared with orthonormal bases $\{\ket{\psi_j}\}_j$ and $\{\ket{\phi_j}\}_j$ of $\HA$. 
Let ${\cal P}:=(\ketbra{p_k}{p_k})_{k=1}^{d^2}$ be a PVM on 
${\cal H}_A\otimes {\cal H}_K$ with an orthonormal basis 
$\{ \ket{p_k}\}_{k=1}^{d^2}$. 

\begin{theorem}\label{QKP_QECC_rev}(Theorem 5 in Ref. \cite{YKMIC15})
If $\ket{\Psi_{\bm \eta}}$ and $\cal P$ provide a solution to the mean king's problem, 
there exists a quantum operation $(L_k)_{k=1}^{d^2}$ on ${\cal H}_K$ and index sets $X^{(J,i)}$ satisfying the following conditions, 
\begin{eqnarray}
&& X^{(J,i)} \cap X^{(J,i')} = \emptyset \hspace{0.2cm} \forall J, \forall i \neq i', \label{Thm_emp_rev}\\
&& M^{(J)}_i = \sum_{k \in X^{(J,i)}} f^{(J,i)}_{k} L_k \hspace{0.2cm} \forall J,i,\label{Thm_decom_rev}\\
&& \bracket{{\mathbb I}\otimes L_k\Psi_{\bm \eta}}{{\mathbb I}\otimes L_{k'}\Psi_{\bm \eta}} = \frac{\alpha}{d} \delta_{kk'} \label{Thm_ortho_rev}
\end{eqnarray}
for some $\alpha > 0$ and $f^{(J,i)}_{k} \in {\mathbb C}$. 
\end{theorem}

From Theorem \ref{QKP_QECC}, 
the initial sate is a code state of a $(d^2,1)$ quantum error-correcting code 
spanned by $\ket{\Psi_{\bm \eta}}$ against the error $({\mathbb I}\otimes L_k)_k$.  
The error satisfies $\sum_kL_k^{\dagger}L_k\leq{\mathbb I}$ when $\alpha=\mathrm{min}\{\eta_j\}_j$ 
from Lemma 4 in the previous work \cite{YKMIC15}. 

\section{Higher-Dimensional Quantum Error-Correcting Codes in Bipartite Systems}\label{Main_sec}
To construct higher-dimensional quantum error-correcting codes 
such that a pair of any code state of the codes and the corresponding measurement provides a solution to the mean king's problem, 
we will utilize Theorem \ref{QKP_QECC} and Theorem \ref{QKP_QECC_rev} effectively. 
Let ${\cal H}_{A'}$ be a quantum system in place of ${\cal H}_A$ satisfying 
$\dim {\cal H}_{A'}=:d_A \geq d=\dim {\cal H}_K$. 
Let 
$$\ket{\Psi_{{\bm \eta},l}}:= \sum_{j=0}^{d-1}\eta_j\ket{\xi_{d (l-1)+j}}\otimes \ket{\phi_j} \in {\cal H}_{A'}\otimes {\cal H}_K, $$
where $l=1,2,\ldots ,
\left\lfloor\frac{d_A}{d}\right\rfloor 
:=\mathrm{max}\{s\in\mathbb Z \mid s\leq \frac{d_A}{d}\}$, 
and $\{\ket{\xi_i}\}_{i=0}^{d_A-1}$ is an orthonormal basis of ${\cal H}_{A'}$. 
Then, we obtain the following theorem. 

\begin{theorem}\label{HQECC_bipartite}
There exists a $(d_Ad, \left\lfloor\frac{d_A}{d}\right\rfloor)$ quantum error-correcting code 
spanned by $\{ \ket{\Psi_{{\bm \eta},l}}\}_{l=1}^{\left\lfloor\frac{d_A}{d}\right\rfloor} 
\subset {\cal H}_{A'}\otimes {\cal H}_K$
such that Alice can guess the king's outcome by using any code state of the code as an initial state 
if the pair of $\ket{\Psi_{\bm \eta}} \in {\cal H}_A\otimes {\cal H}_K$ and ${\cal P}$ provides a solution to the problem. 
\end{theorem}

{\bf Proof.} \
From Theorem \ref{QKP_QECC_rev}, 
if the pair of ${\cal P}$ and $\ket{\Psi_{\bm \eta}} \in {\cal H}_A\otimes {\cal H}_K$ is a solution to the mean king's problem, 
there exist the index sets $X^{(J,i)}$ and the quantum operation $(L_k)_{k=1}^{d^2}$ on ${\cal H}_K$ 
satisfying Eqs. \eqref{Thm_emp_rev}, 
\eqref{Thm_decom_rev}, and \eqref{Thm_ortho_rev}. 

From Eq. \eqref{Thm_ortho_rev}, we observe 
\begin{eqnarray*}
\bracket{{\mathbb I}\otimes L_k\Psi_{{\bm \eta},l}}{{\mathbb I}\otimes L_{k'}\Psi_{{\bm \eta},l'}} &=& 
\sum_{j,j'=0}^{d-1}\eta_j\eta_{j'}\bracket{{\xi_{d (l-1)+j}}}{{\xi_{d (l'-1)+j'}}}\bracket{L_k\phi_j}{L_{k'}\phi_{j'}} \\
&=&  \sum_{j,j'=0}^{d-1}\eta_j\eta_{j'}\delta_{ll'}\delta_{jj'}\bracket{L_k\phi_j}{L_{k'}\phi_{j'}} \\
&=& \frac{\alpha}{d} \delta_{kk'} \delta_{ll'}
\end{eqnarray*}
for the error $({\mathbb I}\otimes L_k)_k$ satisfying $\sum_kL_k^{\dagger}L_k\leq{\mathbb I}$. 
Let $C$ be a $(d_Ad, \left\lfloor\frac{d_A}{d}\right\rfloor)$ quantum code 
spanned by  $\{ \ket{\Psi_{{\bm \eta},l}}\}_{l=1}^{\left\lfloor\frac{d_A}{d}\right\rfloor} 
\subset {\cal H}_{A'}\otimes {\cal H}_K$. 
Then, $C$, $X^{(J,i)}$, and $(L_k)_k$ satisfy Eqs. \eqref{Thm_emp}, \eqref{Thm_decom}, and \eqref{Thm_ortho}. 
Therefore, from Theorem \ref{QKP_QECC}, 
$C$ is the $(d_Ad, \left\lfloor\frac{d_A}{d}\right\rfloor)$ quantum error-correcting code against the error $({\mathbb I}\otimes L_k)_k$ 
and a pair of any code state of $C$ and the corresponding Alice's measurement is a solution to the problem. 
\hfill $\blacksquare$

In the previous work \cite{HHH05}, 
for the problem when king employs measurements described by MUBs, 
the authors showed a solution which consists of a maximal entangled state 
\footnote
{We have this state to substitute 
$\eta_j=1/\sqrt d$ for any $j$ in Eq. \eqref{eq:Sch}. } 
and a PVM measurement constructed from an orthonormal basis 
\footnote{
This basis is listed in the previous work \cite{HHH05}. }. 
Therefore,  from Theorem \ref{HQECC_bipartite}, 
we obtain a higher dimensional quantum error-correcting code from the existence of the solution. 

Here, we give a more specific application example of Theorem \ref{QKP_QECC_rev} in a qubits-system 
to construct a higher dimensional quantum error-correcting code by using Theorem \ref{HQECC_bipartite}. 
This example originates from APPENDIX A in the previous work \cite{YKMIC15}. 
King's measurements are fixed as follows: 
$M^{(1)} := ( M_1^{(1)}:= \ket{+}\bra{+}, M_2^{(1)}:= \ket{-}\bra{-}) , 
M^{(2)} := ( M_1^{(2)}:= \ket{+'}\bra{+'}, M_2^{(2)}:= \ket{-'}\bra{-'}), 
M^{(3)} := ( M_1^{(3)}:= \ket{0}\bra{0}, M_2^{(3)}:= \ket{1}\bra{1})$, 
where $\ket{0} := (1,0)^T,\ket{1} := (0,1)^T, 
\ket{+} := \frac{1}{\sqrt{2}}(1,1)^T, \ket{-} := \frac{1}{\sqrt{2}}(1,-1)^T, 
\ket{+'} := \frac{1}{\sqrt{2}}(1,i)^T$, and $\ket{-'} := \frac{1}{\sqrt{2}}(1,-i)^T$. 
Then, a pair of a Bell state 
$\ket\Psi:=\frac{1}{\sqrt 2}(\ket{00} +\ket{11}) \in 
{\cal H}_A \otimes {\cal H}_K \simeq {\mathbb C}^2\otimes {\mathbb C}^2$ 
and a corresponding PVM measurement constructed from an orthonormal basis 
\footnote{
This basis is listed in the previous work \cite{VAA87}. } 
is a solution to the problem \cite{VAA87}. 
For the solution, from Theorem \ref{QKP_QECC_rev}, 
there exist operators $(L_k)_k$ with $\sum_kL_k^{\dagger}L_k=\mathbb I$ :
\begin{eqnarray}
L_1 := \frac{1}{4} \begin{pmatrix}
                    2    & 1-i \\
                    1+i & 0
                  \end{pmatrix}, &&
L_2 := \frac{1}{4} \begin{pmatrix}
                    2    & -1+i \\
                    -1-i & 0
                  \end{pmatrix}, \nonumber\\
L_3 := \frac{1}{4} \begin{pmatrix}
                    0    & 1+i \\
                    1-i  & 2
                    \end{pmatrix}, &&
L_4 := \frac{1}{4} \begin{pmatrix}
                    0    & -1-i \\
                    -1+i & 2
                    \end{pmatrix} \label{ex_err}
\end{eqnarray}
and also exist index sets listed in Table \ref{table_d2} satisfying Eqs. \eqref{Thm_emp_rev}, 
\eqref{Thm_decom_rev}, and \eqref{Thm_ortho_rev}. 
In particular, 
\begin{eqnarray}
&&M_1^{(1)}=L_1+L_3,\hspace{0.3cm}M_1^{(2)}=L_1+L_4, \hspace{0.3cm} M_1^{(3)}=L_1+L_2,\nonumber\\
&&M_2^{(1)}=L_2+L_4,\hspace{0.3cm}M_2^{(2)}=L_2+L_3, \hspace{0.3cm}M_2^{(3)}=L_3+L_4, \label{ex_obs}
\end{eqnarray} 
and 
\begin{eqnarray*}
\langle {\mathbb I}\otimes L_{k}\Psi |{\mathbb I}\otimes L_{k'}\Psi\rangle =\frac{1}{4}\delta_{kk'}
\end{eqnarray*}
hold 
\footnote{
From Theorem \ref{QKP_QECC}, 
a quantum code spanned by the Bell state is a $(4,1)$ quantum error-correcting code 
against the same error. 
}
. 
From Theorem \ref{HQECC_bipartite}, 
a quantum code spanned by 
$\{ \ket{\Psi_l}:= \frac{1}{\sqrt2} (\ket{\xi_{2(l-1)}}\otimes\ket{0} + \ket{\xi_{2l-1}}\otimes\ket{1}
)\}_{l=1}^{\left\lfloor\frac{d_A}{2}\right\rfloor} \subset {\mathbb C}^{d_A}\otimes {\mathbb C}^2$, 
where $d_A\geq 2$ and $\{\ket{\xi_i}\}_{i=0}^{d_A-1}$ is an orthonormal basis of $\mathbb C^{d_A}$, 
is a $(2d_A, \left\lfloor\frac{d_A}{2}\right\rfloor)$ quantum error-correcting code 
against the error $({\mathbb I}\otimes L_k)_k$ 
such that a pair of any code state of this code and a corresponding Alice's measurement 
is a solution to the problem. 
\begin{small}
\begin{table}[hhhh]
\begin{center}
    \caption{Index sets $X^{(J,i)}$ ($J=1,2,3 : i=1,2$)}
    \begin{tabular}{ccc|ccc|ccc} \hline
    	$J$ & $i$ & $X^{(J,i)}$  & $J$ & $i$ & $X^{(J,i)}$  & $J$ & $i$ & $X^{(J,i)}$\\
    	\hline\hline 
    	$1$ & $1$ & $\{1,3\}$ & $2$ & $1$ & $\{1,4\}$  & $3$ & $1$ & $\{1,2\}$ \\
    	$1$ & $2$ & $\{2,4\}$ & $2$ & $2$ & $\{2,3\}$  & $3$ & $2$ & $\{3,4\}$ \\
    	\hline 
    \end{tabular}
    \label{table_d2}
\end{center}
\end{table}
\end{small}

\section{Higher-Dimensional Quantum Error-Correcting Codes in Multipartite Systems}
In this section, by considering multipartite systems, 
we try to construct a more higher-dimensional quantum error-correcting code. 
Here, we also utilize the previously mentioned quantum operation 
$(L_k)_k$ and index sets $X^{(J,i)}$ satisfying 
Eqs. \eqref{Thm_emp_rev}, \eqref{Thm_decom_rev}, and \eqref{Thm_ortho_rev}
in Theorem \ref{QKP_QECC_rev} 
for the solution $\ket{\Psi_{\bm \eta}}$ and $\cal P$. 

We will start from modifying the setting of the problem as follows. 
Let us consider that Alice can prepare a multipartite system 
${\cal H}^{\otimes n}$ ($\dim{\cal H} =d, n\geq 2$) 
and gives any $l$-th system to king, then she keeps the leftover system in secret. 
Remark that this case is outside the purview of the formulated problems. 
King performs one of the measurements $(M_{i}^{(J)})_{i}$ ($J=1,2,\ldots ,n$) on the $l$-th system. 
After the king's measurement, Alice performs a POVM measurement on the multipartite system ${\cal H}^{\otimes n}$. 
The following state is considered as an initial state: 
\begin{eqnarray}
\ket{\Psi_{{\bm \eta},i_1,\ldots ,i_n}}:=\sum_{j=0}^{d-1}\eta_{j}
\ket{\phi_{j\oplus i_1}}\otimes\ket{\phi_{j\oplus i_2}}\otimes
\cdots\otimes\ket{\phi_{j\oplus i_n}}, \label{gGHZ}
\end{eqnarray}
where $i_u\in\{ 0,1,\ldots ,d-1\}$ and $j\oplus i_u := j+ i_u$ mod $d$. 
This state is inspired by generalized Greenberger-Horne-Zeilinger (GHZ) states \cite{BB01,KB02}. 

\begin{lemma}\label{m_Inn}
\begin{eqnarray}
\bracket{\tilde L_k\Psi_{{\bm \eta},i_1,\ldots ,i_n}}{\tilde L_{k'}\Psi_{{\bm \eta},i_1,\ldots ,i_n}} = \frac{\alpha}{d}\delta_{kk'} \label{m_ip}
\end{eqnarray}
holds, where $\tilde L_k:=\mathbb{I}\otimes\cdots\otimes\mathbb{I}\otimes L_k\otimes\mathbb{I}\otimes\cdots\otimes\mathbb{I}$. 
And 
\begin{eqnarray}
\bracket{\tilde L_k\Psi_{{\bm \eta},i_1,\ldots ,i_n}}{\tilde L_{k'}\Psi_{{\bm \eta},i'_1,\ldots ,i'_n}} = 0 \label{m_ortho}
\end{eqnarray}
holds if there exist $s$ and $t$ ($s\neq t$ and $s,t\neq l$) such that $i_s\ominus i'_s \neq i_t\ominus i'_t $, 
where $i_u\ominus i'_u := i_u-i'_u$ $\rm{mod}$ $d$. 
\end{lemma}

{\bf Proof.} \
We observe 
\begin{eqnarray}
\bracket{\tilde L_k\Psi_{{\bm \eta},i_1,\ldots ,i_n}}{\tilde L_{k'}\Psi_{{\bm \eta},i'_1,\ldots ,i'_n}}  
= \sum_{j,j'}\eta_{j}\eta_{j'}\bracket{\phi_{j\oplus i_0}}{\phi_{j'\oplus i'_0}}\cdots \bracket{L_k\phi_{j\oplus i_l}}{L_{k'}\phi_{j'\oplus i'_l}}\cdots \bracket{\phi_{j\oplus i_{n-1}}}{\phi_{j'\oplus i'_{n-1}}}. \label{expand}
\end{eqnarray}
From Eq. \eqref{Thm_ortho_rev}, the right hand side of Eq. \eqref{expand} 
$= \sum_{j}\eta_{j}^2 \bracket{L_k\phi_{j\oplus i_l}}{L_{k'}\phi_{j\oplus i_{l}}} 
= \sum_{j}\eta_{j}^2 \bracket{L_k\phi_{j}}{L_{k'}\phi_{j}} 
= \frac{\alpha}{d}\delta_{kk'}$ when $i_s=i'_s$ for any $s$. 
This implies that Eq. \eqref{m_ip} holds. 

We show that $i_s\ominus i'_s = i_t\ominus i'_t $ holds 
if there exist $j$ and $j'$ such that $\bracket{\phi_{j\oplus i_s}}{\phi_{j'\oplus i'_s}}\bracket{\phi_{j\oplus i_t}}{\phi_{j'\oplus i'_t}}\neq 0$ 
for fixed $s$ and $t$ ($s\neq t$ and $s,t\neq l$). 
Since $\{ \ket{\phi_j}\}_j$ is the orthonormal basis, 
$j\oplus i_s=j'\oplus i'_s$ and $j\oplus i_t=j'\oplus i'_t$ holds 
if $\bracket{\phi_{j\oplus i_s}}{\phi_{j'\oplus i'_s}}\bracket{\phi_{j\oplus i_t}}{\phi_{j'\oplus i'_t}}\neq 0$ holds. 
This implies that $(j+i_s)-(j'+i'_s) = a_{ss'}d$ ($\Leftrightarrow i_s-i'_s=a_{ss'}d+j'-j$) 
and $(j+i_t)-(j'+i'_t) = a_{tt'}d$ ($\Leftrightarrow i_t-i'_t=a_{tt'}d+j'-j$) 
hold for some integers $a_{ss'}$ and $a_{tt'}$. 
Then $i_s\ominus i'_s = i_t\ominus i'_t$ holds. 
Therefore, if there exist $s$ and $t$ such that $i_s\ominus i'_s \neq i_t\ominus i'_t$, 
$\bracket{\phi_{j\oplus i_s}}{\phi_{j'\oplus i'_s}}\bracket{\phi_{j\oplus i_t}}{\phi_{j'\oplus i'_t}}= 0$ for any $j$ and $j'$ 
which implies that Eq. \eqref{expand} is equal to $0$. 
\hfill $\blacksquare$

\begin{theorem}\label{m_main}
There exists a $(d^n, g(\geq 2))$ quantum error-correcting code constructed from the states described by Eq. \eqref{gGHZ} 
such that Alice can guess the king's outcome by using any code state of the code 
as an initial state under the above setting 
if the pair of $\ket{\Psi_{\bm \eta}} \in {\cal H}_A\otimes {\cal H}_K$ and ${\cal P}$ provides a solution to the problem. 
\end{theorem}

{\bf Proof.} \
From Theorem \ref{QKP_QECC_rev}, 
there exist $(L_k)_k$ and $X^{(J,i)}$ satisfy Eqs. \eqref{Thm_emp_rev}, 
\eqref{Thm_decom_rev}, and \eqref{Thm_ortho_rev} 
if the pair of $\ket{\Psi_{\bm \eta}}$ and ${\cal P}$ provides a solution. 
We also observe 
\begin{eqnarray}
\tilde{M}_i^{(J)}=\sum_{k\in X^{(J,i)}}\tilde{L}_k, \hspace{0.2cm}
X^{(J,i)}\cap X^{(J,i')}=\emptyset ,\hspace{0.2cm}\mathrm{and}\hspace{0.2cm} 
\sum_{k}\tilde{L}_k^\dagger \tilde{L}_k  \leq \mathbb I ,\label{decom}
\end{eqnarray}
where $\tilde{M}_i^{(J)}:=\mathbb{I}\otimes\cdots\otimes\mathbb{I}\otimes M_i^{(J)}\otimes\mathbb{I}\otimes\cdots\otimes\mathbb{I}$. 

Let $\tilde S$ be a set of the states described by Eq. \eqref{gGHZ} such that 
$\bracket{\tilde L_k\Psi_{{\bm \eta},i_1,\ldots ,i_n}}{\tilde L_{k'}\Psi_{{\bm \eta},i'_1,\ldots ,i'_n}} = 0$ holds 
for any pair of different states in $\tilde S$ 
and let $g$ be the number of the elements of $\tilde S$. 
By observing the inner product, 
we find that Eq. \eqref{m_ip} holds for any state in $\tilde S$ 
and $g\geq 2$ from Lemma \ref{m_Inn}. 
Let $\tilde C$ be a $(d^n,g)$ quantum code spanned by $\tilde S$ and $\tilde P$ the projection onto $\tilde C$. 
Then, 
\begin{eqnarray}
\tilde P\tilde L_k^{\dagger}\tilde L_{k'}\tilde P=\frac{\alpha}{d}\delta_{kk'}\tilde P \label{ortho}
\end{eqnarray}
holds. 

We regard the $l$-th system and the leftover system in the same light as ${\cal H}_K$ and ${\cal H}_A$, respectively. 
Then, $(\tilde L_k)_k$ and $X^{(J,i)}$ satisfy Eqs. \eqref{Thm_emp}, \eqref{Thm_decom}, and \eqref{Thm_ortho} for $\tilde C$ from Eqs. \eqref{ortho} and \eqref{decom}. 
Therefore, from Theorem \ref{QKP_QECC}, 
it is straightforward that $\tilde C$ is a $(d^n, g)$ quantum error-correcting code 
and a pair of any code state of $\tilde C$ and a corresponding Alice's measurement are a solution to the problem. 
\hfill $\blacksquare$

We give a $(2^n, 2^{n-2})$ quantum error-correcting code constructed from GHZ states 
as an application example of Theorem \ref{m_main} in a multipartite qubits-system. 
Let ${\cal H}^{\otimes n}\simeq ({\mathbb C}^2)^{\otimes n}$ be a multipartite qubits-system prepared by Alice, 
$({M}_i^{(J)})_{i=1,2}$ ($J=1,2,3$) described by Eq. \eqref{ex_obs} 
be the measurements employed by King, 
$({L}_k)_{k=1}^4$ defined as Eq. \eqref{ex_err} and $X^{(J,i)}$ listed in Table \ref{table_d2} 
be the corresponding quantum operation and the index sets, respectively. 
Then, $\tilde{M}_i^{(J)}=\sum_{k\in X^{(J,i)}}\tilde{L}_k, X^{(J,i)}\cap X^{(J,i')}=\emptyset$, 
and $\sum_k\tilde{L}_k^\dagger\tilde{L}_k=\mathbb I$ hold. 

We will try to construct a quantum code spanned by GHZ states
\footnote{
We have this state to substitute 
$d=2, \eta_{j}=1/\sqrt 2, \ket{\phi_{j\oplus i_u}}=\ket{j\oplus i_u}$ in Eq. \eqref{gGHZ}. 
} 
defined as 
$$
\ket{\Psi_{i_1,\ldots ,i_n}} := 
\frac{1}{\sqrt 2}(\ket{i_1}\otimes\ket{i_2}\otimes\cdots\otimes\ket{i_n}+
\ket{\bar{i_1}}\otimes\ket{\bar{i_2}}\otimes\cdots\otimes\ket{\bar{i_n}}), $$
where $i_u \in \{ 0,1 \}$ and $\ket{\bar{i_u}}:=\ket{i_u\oplus 1}$. 
Then, 
$\bracket{\tilde L_k\Psi_{i_1,\ldots ,i_n}}{\tilde L_{k'}\Psi_{i_1,\ldots ,i_n}}=\frac 1 4 \delta_{kk'}$ holds. 
Let $\tilde S_{ghz}$ be a set of the GHZ states such that 
$\bracket{\tilde L_k\Psi_{i_1,\ldots ,i_n}}{\tilde L_{k'}\Psi_{i'_1,\ldots ,i'_n}}=0$
for any pair of different states in $\tilde S_{ghz}$. 
By observing the inner product, 
we find that the number of the elements of $\tilde S_{ghz}$ is $2^{n-2}$. 
Let $\tilde C_{ghz}$ be a $(2^n, 2^{n-2})$ quantum code spanned by $\tilde S_{ghz}$ and $\tilde P_{ghz}$ the projection onto $\tilde C_{ghz}$. 
Then, 
$\tilde P_{ghz}\tilde L_k^{\dagger}\tilde L_{k'}\tilde P_{ghz}=\frac 1 4 \delta_{kk'}\tilde P_{ghz}$ 
holds. 
In the same way as the proof of Theorem \ref{m_main}, 
it is straightforward that $\tilde C_{ghz}$ is a $(2^n, 2^{n-2})$ quantum error-correcting code 
and a pair of any code state of $\tilde C_{ghz}$ and a corresponding Alice's measurement is a solution 
from Theorem \ref{QKP_QECC}. 

\section{Summary}
The solution to the mean king's problem by using quantum error-correcting codes has been shown 
in the previous work. 
In the solution, Alice can guess the king's outcomes correctly 
when she utilize any code state of the code  as an initial state 
and a measurement to discriminate error corresponding to the king's measurement. 
However, the general construction of such codes under any problem setting is not known. 
In this paper, we showed the constructions of higher dimensional quantum error-correcting codes 
based on the solution $\ket{\Psi_{\bm \eta}}$ and ${\cal P}$ 
in both of the bipartite systems and the multipartite systems. 
The dimension of our constructed codes is greater than or equal to $2$ 
and it is recalled that a code state is defined as a pure state of a quantum code. 
This implies that we can find more large solution space in the context of the solution using quantum error-correcting codes 
if $\ket{\Psi_{\bm \eta}}$ and ${\cal P}$ provide a solution to the mean king's problem.

\end{document}